# Luminosity and Physics Considerations on HL-LHC and HE-LHC based *μp* Colliders

U. Kaya[a], B. Ketenoglu[b,*], S. Sultansoy[a,c], F. Zimmermann[d]

[a]*TOBB University of Economics and Technology, Ankara, Turkey*
[b]*Department of Engineering Physics, Ankara University, Ankara, Turkey*
[c]*ANAS Institute of Physics, Baku, Azerbaijan*
[d]*CERN, Geneva, Switzerland*

**Abstract**

Construction of future Muon Collider tangential to the Large Hadron Collider will give opportunity to realize *μp* collisions at multi-TeV center of mass energies. Using the nominal parameters of high luminosity and high energy upgrades of the LHC, as well as the design parameters of muon colliders, it is shown that $L_{\mu p}$ of order of $10^{33}$ $cm^{-2}s^{-1}$ is achievable for different options with $\sqrt{s_{\mu p}}$ from 4.58 TeV to 12.7 TeV. Certainly, proposed *μp* colliders have a huge potential for clarifying QCD basics and searches for new physics.

## 1. Introduction

It is known that lepton-hadron collisions played a crucial role in our understanding of matter's structure (proton form-factors, quark-parton model, EMC effect and so on). The first electron-proton collider HERA explored this structure further and provided parton distribution functions (PDFs) for the LHC and Tevatron. Construction of a TeV (or multi-TeV) scale lepton-hadron collider is mandatory both for clarifying of QCD basics, which is responsible for 98% of mass of the visible part of our Universe, and to provide PDFs for adequate interpretation of incoming data from HL/HE-LHC [1, 2] and FCC/SppC [3, 4].

On the other hand, while the electro-weak part of the Standard Model (SM) has been completed with discovery of Higgs boson at the LHC, this is not the case for QCD part: the confinement hypothesis should be clarified experimentally. In this respect, energy frontier lepton-hadron colliders will play a role analogous to the LHC, which clarify the Higgs mechanism hypothesis.

Today, linac-ring type *ep* colliders are considered as sole realistic way to (multi-) TeV scale in lepton-hadron collisions (see the review [5] and references therein) and LHeC [6] is the most promising candidate. However, situation may change in the coming years: *μp* colliders can come forward depending on progress in muon production and cooling issues.

TeV energy muon-proton colliders [7, 8] were proposed two decades ago as alternatives to linac-HERA and linac-LHC

*Corresponding author: bketen@eng.ankara.edu.tr

based *ep*/*γp* colliders (see review [9] and references therein). Two years later an ultimate √s=100 TeV *μp* collider (with additional 50 TeV proton ring in √s=100 TeV muon collider tunnel) was suggested in [10]. It should be noted that luminosities of *μp* collisions in [7] and [8] were over-estimated (see subsections 3.2 and 2.3 in [10]).

Recently, FCC and SppC based energy frontier muon-hadron colliders have been proposed in [11] and [12], respectively. In order to complete picture and keeping in mind the approved high luminosity and proposed high energy upgrades of the LHC (HL-LHC and HE-LHC, respectively), corresponding *μp* colliders have to be considered as well.

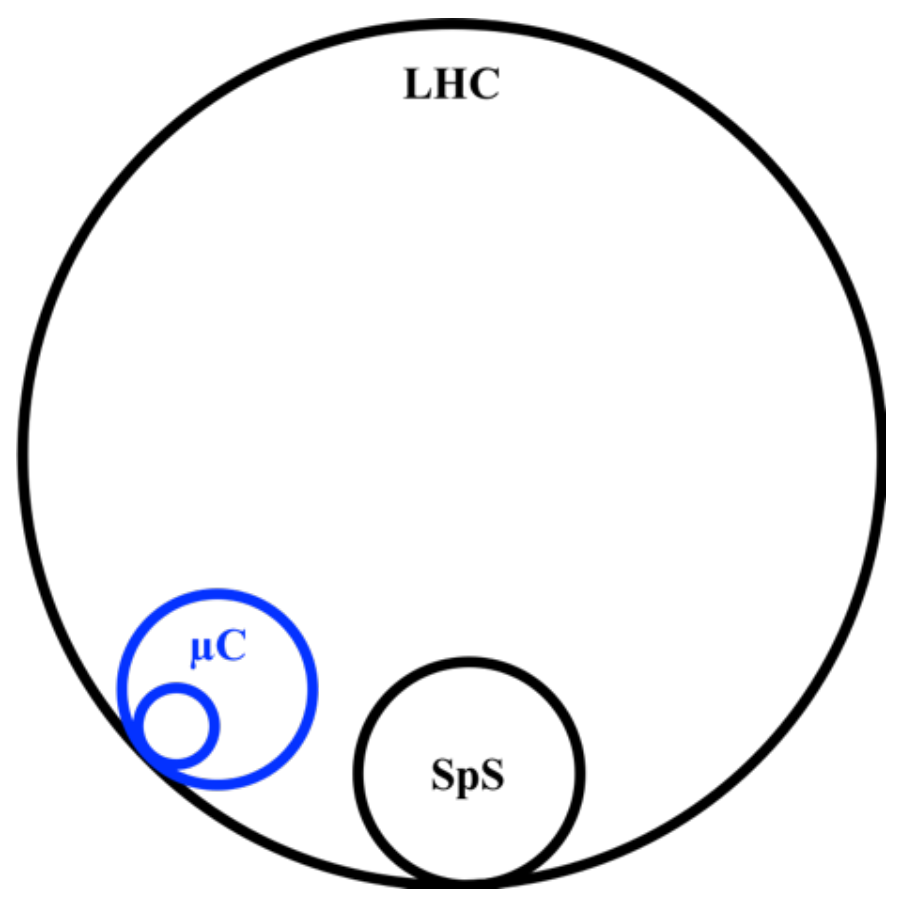

Figure 1. Schematic drawing of proposed *μp* colliders

In this paper we propose the construction of Muon Colliders (MC) tangential to the LHC (see Fig. 1) to allow for multi-TeV center-of-mass energy *μp* collisions. Nominal parameters of HL-LHC, HE-LHC and MC are presented in Section 2. Center of mass energies and luminosities for different *μp* collider options are estimated in Section 3. Then, in Section 4 we briefly discuss physics at these colliders. Finally, in Section 5 we present our conclusions and recommendations.

## 2. Parameters of HL-LHC, HE-LHC and Muon Colliders

In this section, we present parameters of HL-LHC, HE-LHC and MC, which are used for estimation of main parameters of *μp* colliders in the following section. Table 1 presents nominal parameters for HL-LHC and HE-LHC [13]. Parameters of proton ring upgraded for ERL60 related *ep* colliders are given in Table 2.

Table 1. HL-LHC and HE-LHC design parameters [13]

| **Parameter [Unit]** | **HL-LHC** | **HE-LHC** |
|---|---|---|
| √s energy [TeV] | 14 | 27 |
| Circumference [km] | 26.7 | 26.7 |
| Beam current [A] | 1.12 | 1.12 |
| Bunch population [$10^{11}$] | 2.2 | 2.2 |
| Number of bunches per beam | 2760 | 2808 |
| RMS bunch length [mm] | 90 | 90 |
| Bunch spacing [ns] | 25 | 25 |
| Normalized rms emittances [μm] | 2.5 | 2.5 |
| $\beta_{x,y}$ @ IP [m] | 0.15 | 0.45 |
| RMS beam sizes [μm] | 7.1 | 9 |
| Peak luminosity per IP [$cm^{-2}s^{-1}$] | $5x10^{34}$ | $16x10^{34}$ |

Table 2. ERL60 ⊗ (HL-LHC & HE-LHC) *ep* collider parameters [13]

| **Parameter [Unit]** | **HL-LHC** | **HE-LHC** |
|---|---|---|
| $E_p$ [TeV] | 7 | 13.5 |
| $E_e$ [GeV] | 60 | 60 |
| $\sqrt{s_{ep}}$ energy [TeV] | 1.3 | 1.7 |
| Bunch spacing [ns] | 25 | 25 |
| Protons per bunch [$10^{11}$] | 2.2 | 2.5 |
| Proton normalized emittances [µm] | 2 | 2.5 |
| Electrons per bunch [$10^9$] | 2.3 | 3 |
| Electron current [mA] | 15 | 20 |
| Protons $\beta$ @ IP [cm] | 7 | 10 |
| Proton beam sizes [µm] | 4.45 | 4.17 |
| Luminosity [$cm^{-2}s^{-1}$] | $8\times10^{33}$ | $12\times10^{33}$ |

Main parameters of TeV energy muon collider options [14] are listed Table 3.

Table 3. Muon Collider parameters [14]

| **Parameter [Unit]** | **Multi-TeV** | | |
|---|---|---|---|
| $\sqrt{s}$ energy [TeV] | 1.5 | 3 | 6 |
| Circumference [km] | 2.5 | 4.5 | 6 |
| Bunch population [$10^{12}$] | 2 | 2 | 2 |
| Bunch length [cm] | 1 | 0.5 | 0.2 |
| Normalized emittances [µm] | 25 | 25 | 25 |
| $\beta_{x,y}$ @ IP [cm] | 1 | 0.5 | 0.25 |
| Beam sizes [µm] | 5.9 | 2.95 | 1.48 |
| Repetition rate [Hz] | 15 | 12 | 6 |
| Number of IPs | 2 | 2 | 2 |
| Average luminosity [$cm^{-2}s^{-1}$] | $1.25\times10^{34}$ | $4.4\times10^{34}$ | $12\times10^{34}$ |

## 3. Main parameters of $\mu p$ colliders

The general expression for luminosity of the LHC-based $\mu p$ colliders is given by:

$$L_{\mu p} = \frac{N_\mu N_p}{4\pi \, max\left[\sigma_{x_p}, \sigma_{x_\mu}\right] max\left[\sigma_{y_p}, \sigma_{y_\mu}\right]} min\left[f_{c_p}, f_{c_\mu}\right] \quad (1)$$

where $N_\mu$ and $N_p$ are numbers of muons and protons per bunch, respectively; $\sigma_{x_p}$ ($\sigma_{x_\mu}$) and $\sigma_{y_p}$ ($\sigma_{y_\mu}$) are the horizontal and vertical proton (muon) beam sizes at interaction point (IP); $f_{c_\mu}$ and $f_{c_p}$ are MC and LHC bunch frequencies. $f_c$ is expressed by $f_c = N_b f_{rev}$, where $N_b$ denotes the number of bunches and $f_{rev}$ means revolution frequency. Some of these parameters can be rearranged in order to maximize $L_{\mu p}$ but one should note that there are main limitations due to beam-beam effects that should be kept in mind. Beam-beam parameter for proton beam is given by:

$$\xi_{x_p} = \frac{N_\mu r_p \beta_p^*}{2\pi\gamma_p \sigma_{x_\mu}(\sigma_{x_\mu}+\sigma_{y_\mu})},$$

$$\xi_{y_p} = \frac{N_\mu r_p \beta_p^*}{2\pi\gamma_p \sigma_{y_\mu}(\sigma_{y_\mu}+\sigma_{x_\mu})},$$

where $r_p$ is classical radius of proton, $r_p$=1.54x$10^{-18}$ m, $\beta_p^*$ is beta function of proton beam at IP, and $\gamma_p$ is the Lorentz factor of the proton beam. $\sigma_{x_\mu}$ and $\sigma_{y_\mu}$ are horizontal and vertical sizes of muon beam at IP, respectively. The beam-beam parameter for the muon beam is given by:

$$\xi_{x_\mu} = \frac{N_p r_\mu \beta_\mu^*}{2\pi\gamma_\mu \sigma_{x_p}(\sigma_{x_p}+\sigma_{y_p})}, \quad (2)$$

$$\xi_{y_\mu} = \frac{N_p r_\mu \beta_\mu^*}{2\pi\gamma_\mu \sigma_{y_p}(\sigma_{y_p}+\sigma_{x_p})}, \quad (3)$$

where $r_\mu$=1.37x10$^{-17}$ m is classical muon radius, $\beta_\mu^*$ is beta function of muon beam at IP, and $\gamma_\mu$ is the Lorentz factor of muon beam. $\sigma_{x_p}$ and $\sigma_{y_p}$ are horizontal and vertical sizes of proton beam at IP, respectively.

Keeping in mind that both LHC and MC have round beams, luminosity equations for $\mu\mu$ and $pp$ are given by:

$$L_{\mu\mu} = \frac{N_\mu^2}{4\pi\sigma_\mu^2} f_{c\mu} \quad (4)$$

$$L_{pp} = \frac{N_p^2}{4\pi\sigma_p^2} f_{cp} \quad (5)$$

Concerning muon-proton collisions, according to Eq.1, one should use the larger of the two transverse beam sizes and the smaller of the two collision frequency values. Keeping in mind that $f_{c\mu}$ is smaller than $f_{cp}$ by more than two orders, the following relation between $\mu p$ and $\mu\mu$ luminosities is found for round beams:

$$L_{\mu p} = \left(\frac{N_p}{N_\mu}\right)\left(\frac{\sigma_\mu}{max[\sigma_p, \sigma_\mu]}\right)^2 L_{\mu\mu} \quad (6)$$

Using nominal (upgraded) parameters of HL-LHC and MC from Tables 1, 2 and 3, parameters of HL-LHC based $\mu p$ colliders are estimated according to Eq. 6 and presented in Table 4.

Table 4. Center of mass energies and luminosities of HL-LHC based $\mu p$ colliders

| $E_\mu$, TeV | √s, TeV | L (nominal), $10^{33}$ cm$^{-2}$s$^{-1}$ | L (upgraded), $10^{33}$ cm$^{-2}$s$^{-1}$ |
|---|---|---|---|
| 0.75 | 4.58 | 0.95 | 1.4 |
| 1.5 | 6.48 | 0.84 | 2.1 |
| 3 | 9.16 | 0.57 | 1.5 |

Center of mass energies and luminosity values for HE-LHC based $\mu p$ colliders, evaluated in the same way, are given in Table 5.

Table 5. Center of mass energies and luminosities of HE-LHC based $\mu p$ colliders

| $E_\mu$, TeV | √s, TeV | L (nominal), $10^{33}$ cm$^{-2}$s$^{-1}$ | L (upgraded), $10^{33}$ cm$^{-2}$s$^{-1}$ |
|---|---|---|---|
| 0.75 | 6.36 | 0.59 | 1.6 |
| 1.5 | 9 | 0.52 | 2.8 |
| 3 | 12.7 | 0.36 | 1.9 |

Concerning beam tune shifts, for round beams Eqs. (2) and (3) turn into:

$$\xi_p = \frac{N_\mu r_p \beta_p^*}{4\pi\gamma_p \sigma_\mu^2} \quad (7)$$

$$\xi_\mu = \frac{N_p r_\mu \beta_\mu^*}{4\pi\gamma_\mu \sigma_p^2} \quad (8)$$

Matching transverse sizes of proton and muon beams results in:

$$\xi_p = \frac{N_\mu r_p}{4\pi\varepsilon_N^p} \quad (9)$$

$$\xi_\mu = \frac{N_p r_\mu}{4\pi\varepsilon_N^\mu} \quad (10)$$

Putting corresponding parameters from Tables 1-3 into Eqs. 9-10, we obtain the tune shift values given in Table 6.

Table 6. Beam-beam tune shifts

| | | $\xi_p$ | $\xi_\mu$ |
|---|---|---|---|
| HL-LHC | Nominal | 0.098 | 0.0096 |
| | Upgraded | 0.12 | 0.0096 |
| HE-LHC | Nominal | 0.098 | 0.0096 |
| | Upgraded | 0.098 | 0.011 |

One can see from Table 6 that while $\xi_\mu$ values are acceptable, $\xi_p$ values should be reduced by an order. According to Eq. 9, this can be handled by decrease of $N_\mu$ and/or increase of normalized emittance of proton beam, resulting in corresponding decrease of luminosity.

Several years ago, the AloHEP software [15, 16] has been developed to estimate parameters for linac-ring type *ep* colliders such as luminosity, beam-beam tune shift etc. Recently, the software AloHEP has been extended to cover all types of colliding beams and collision schemes [17]. Using AloHEP, we have double-checked the parameters given in Tables 4, 5 and 6, resulting in a well-consistent outcome.

## 4. Physics search potential

Because of high center of mass energy and luminosity values, the LHC based $\mu p$ colliders have a huge potential for the SM and BSM searches. Concerning SM physics, they will provide precision PDFs for the HE-LHC, FCC and SppC. Small $x$ Bjorken region, which is crucial for understanding of QCD basics, can be explored down to $10^{-8}$. Precision Higgs physics is another important topic, which should be analyzed in detail. In Table 7, we present the achievable $x$ Bjorken values for different options of the LHC based $\mu p$ colliders. Let us mention that $Q^2$=25 GeV$^2$ corresponds to perturbative QCD, whereas $x_B<10^{-6}$ implies high parton densities.

Table 7. Achievable $x$ Bjorken values at the LHC based μp colliders.

| √s, TeV | $Q^2$= 1 GeV$^2$ | $Q^2$= 25 GeV$^2$ |
|---|---|---|
| 4.58 | $4.8 \times 10^{-8}$ | $1.2 \times 10^{-6}$ |
| ~ 6.4 | $2.5 \times 10^{-8}$ | $6.1 \times 10^{-7}$ |
| ~ 9 | $1.3 \times 10^{-8}$ | $3.1 \times 10^{-7}$ |
| 12.7 | $6.2 \times 10^{-9}$ | $1.6 \times 10^{-7}$ |

Concerning BSM physics, LHC based $\mu$p colliders are comparable or essentially exceed the potential of the LHC itself in a lot of topics, such as leptoquarks related to the second family leptons, excited muon, excited muon neutrino, color octet muon, contact interactions, SUSY, RPV SUSY (especially resonant production of corresponding squarks), extended gauge symmetry etc.

As an example, we present discovery limits for color octet muon (see [18] and references therein) at the HL-LHC, HE-LHC, MC with √s=6 TeV and corresponding $\mu p$ colliders in Fig. 2 assuming the compositeness scale to be equal to color octet muon mass. Here, pair production at the LHC and MC and resonant production at $\mu p$ colliders have been considered.

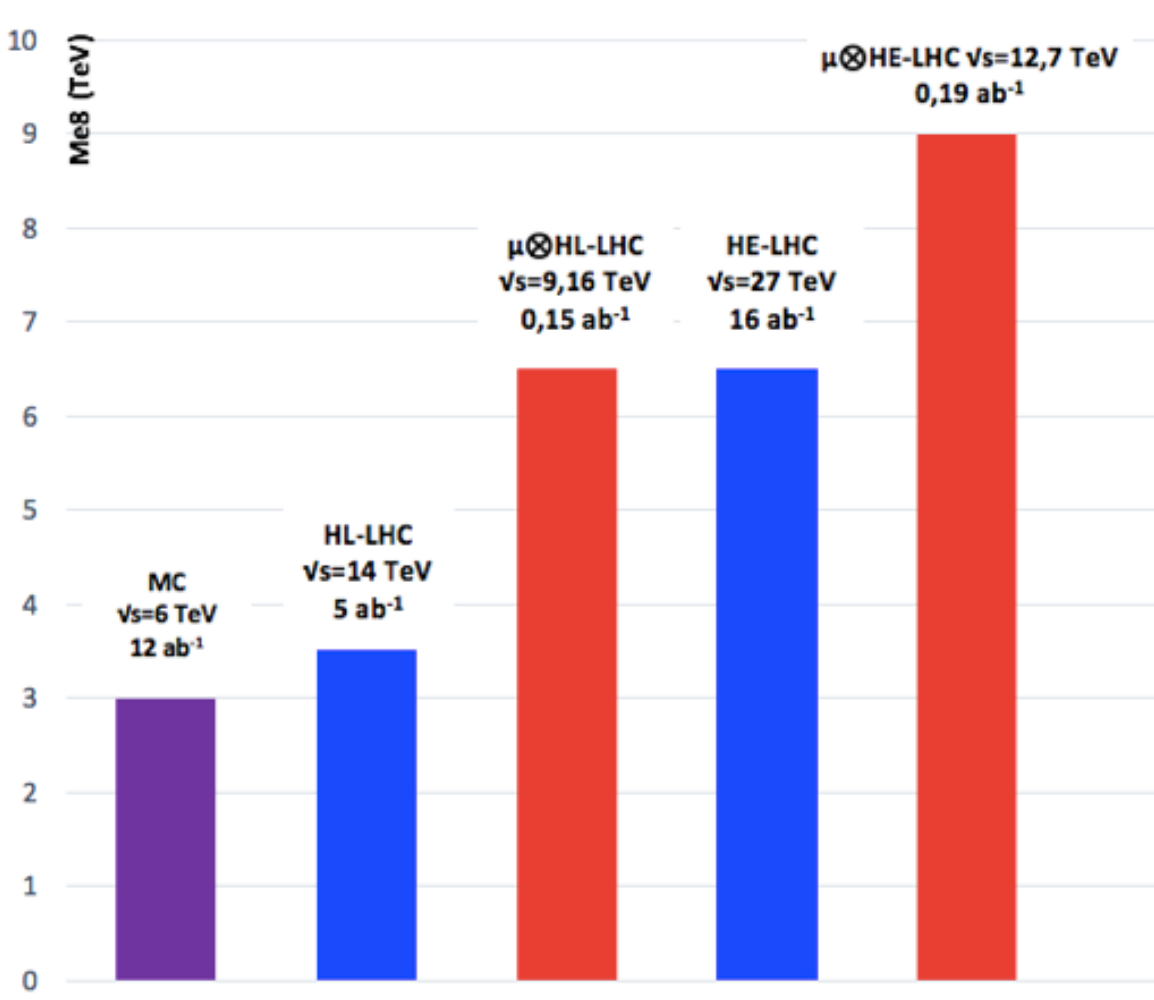

Figure 2. Discovery limits for Color Octet Muon at the MC with √s=6 TeV, HL-LHC, HE-LHC and corresponding *μp* colliders

Let us mention that importance of *μp* colliders is emphasized by results of muon g-2 factor measurement (4.2$\sigma$ discrepancy with the theoretical calculations [19]) and lepton flavour violation in B-meson decays (4.0$\sigma$ discrepancy [20]), which could be a manifestation of the muon-related new physics. Finally, the potential of the LHC based *μp* collider (with $E_\mu$=1 TeV, $E_p$=7 TeV and $\sqrt{s}$=5.3 TeV) in measuring the Higgs properties, probing the R-parity-violating supersymmetry and testing heavy new physics in the muon g−2 anomaly has been considered in a recent paper [21].

## 5. Conclusion

It is shown that the construction of future muon collider (or dedicated *μ*-ring) tangential to the LHC will give the opportunity to realize *μp* colliders with multi-TeV center-of-mass energies at a luminosity of order of $10^{33}$ $cm^{-2}s^{-1}$. Obviously, such colliders will essentially enlarge the physics search potential of the HL/HE-LHC for both the SM and BSM phenomena. Therefore, systematic studies of accelerator, detector and physics search aspects of the LHC based *μp* colliders are necessary for long-term planning in the field of High Energy Physics.

Finally, one can consider two-stage scenario for the LHC based lepton-hadron colliders: the LHeC option with 9 km e-ring [22] as the first stage, already tangential to the LHC, followed by the construction of *μ*-ring in the same tunnel.